\documentclass[11pt]{article}
\usepackage[utf8]{inputenc}
\usepackage{jheppub}
\usepackage{amsmath}
\usepackage{physics}
\usepackage{tikz}
\usepackage{cleveref}
\usepackage{comment}
\usepackage{natbib}

\usepackage{subfigure}

\newcommand{\ext}{\mathrm{d}}
\newcommand{\AdS}{\mathrm{AdS}}
\newcommand{\be}{\begin{equation}}
\newcommand{\ee}{\end{equation}}

\title{Brane dynamics from the first law of entanglement}
\author[a]{Sean Cooper,}
\emailAdd{seancooper@phas.ubc.ca}
\author[a,b]{Dominik Neuenfeld,}
\emailAdd{dneuenfeld@perimeterinstitute.ca}
\author[a]{Moshe Rozali,}
\emailAdd{rozali@phas.ubc.ca}
\author[a]{David Wakeham}
\emailAdd{david.a.wakeham@gmail.com}
\date{\today}

\affiliation[a]{Department of Physics and Astronomy,\\
University of British Columbia, 6224 Agricultural Road, 
Vancouver,
BC V6T 0C2, Canada}
\affiliation[b]{Perimeter Institute for Theoretical Physics, \\
31 Caroline Street N., Waterloo, Ontario N2L 2Y5, Canada
}

\abstract{
In this note, we study the first law of entanglement in a boundary conformal field theory (BCFT) dual to warped AdS cut off by a brane.
Exploiting the symmetry of boundary-centered half-balls in the BCFT, and using Wald's covariant phase space formalism in the presence of boundaries, we derive constraints from the first law for a broad range of covariant bulk Lagrangians.
We explicitly evaluate these constraints for Einstein gravity, and find a local equation on the brane which is precisely the Neumann condition of Takayanagi [arXiv:1105.5165] at linear order in metric perturbations. This is analogous to the derivation of Einstein's equations from the first law of entanglement entropy.
This machinery should generalize to give local linearized equations of motion for higher-derivative bulk gravity with additional fields.
}

\begin{document}

\maketitle

\makeatletter
\@starttoc{toc}
\makeatother

\newpage

\section{Introduction}

Quantum field theories at critical points, i.e., at fixed points of renormalization group flow, are described by \emph{conformal field theories} (CFTs).
A quantum field theory with boundary, whose bulk degrees of freedom and boundary condition are both critical, is described by a \emph{boundary CFT} (BCFT) \cite{Cardy1984}. In the condensed matter context, these theories describe the critical dynamics of systems with defects.

At large $N$, the holographic correspondence gives a dual description of CFTs in terms of semiclassical gravity in asymptotically Anti-de-Sitter (AdS) spaces \cite{maldacena1999large,Gubser1998, Witten1998a}.
A particular holographic ansatz for BCFTs at large $N$ and strong coupling, called the \emph{AdS/BCFT correspondence}, was proposed in \cite{Karch2000,Takayanagi2011, Fujita2011}.
Loosely speaking, it states that a CFT boundary is dual to an \emph{end-of-the-world }(ETW) brane obeying Neumann boundary conditions.
Recently, holographic BCFTs have found applications in the construction of explicit black hole microstates \cite{Almheiri2018, Cooper2018}, the possible resolution of the black hole information problem \cite{Almheiri2019a, Almheiri2019TheHole,Almheiri:2019yqk,Rozali:2019day}, emergent spacetime \cite{VanRaamsdonk2018}, holographic duals of quenches \cite{Shimaji2018}, and even quantum cosmology \cite{Cooper2018, Antonini2019}. These applications have largely used the ETW prescription for the holographic dual of a BCFT.

In some top-down supergravity models, the gravity dual of a BCFT can be described explicitly \cite{Chiodaroli2011, Chiodaroli2012a, Chiodaroli2012}.
The nontrivial role of internal warping and additional supergravity fields suggests that the ETW prescription does not always capture the full details of the dual geometry. 
It remains possible, however, that a large class of BCFTs can be effectively described by these ETW branes, similar to the regular bottom-up approach, where the AdS/CFT correspondence can model strong coupling dynamics without necessarily embedding the holographic dual in a UV complete theory of gravity.

We may then wonder if the ETW prescription for a holographic BCFT is consistent and, if the answer is affirmative, whether there is a procedure for deriving consistent bulk dynamics from the BCFT. For a boundary-less CFT, such a procedure was given in \cite{Faulkner2014} where it was shown that under certain conditions one can derive the Einstein's equations at linear order around pure AdS from the first law of entanglement entropy.\footnote{See \cite{Faulkner:2017tkh} for an extension to second order.}

The aim of this note is to demonstrate that the first law of entanglement can similarly be used in the context of holographic BCFTs to derive the brane equations of motion, carefully employing the covariant phase space formalism \cite{Crnkovic:1987tz, Lee1990,Iyer1994,Wald1999, Harlow2019}.
This is a nontrivial consistency check for the ETW prescription and points to a systematic procedure to derive the dual for more general situations.

If the linearized Einstein equations in the bulk hold, we will see that the first law requires a certain form $\chi_{A}$, associated with boundary-centred half-balls $A$ in the BCFT, to vanish when integrated over a corresponding region $\mathcal{B}_A$ of the brane:
\[
\int_{\mathcal{B}_A} \chi_A = 0.
\]

We will show that, for Einstein gravity, if the background obeys a Neumann condition, then we can turn the global constraints on $\chi_A$ into a local constraint:
\[
\mathcal{N}_{\mu\nu} = 0 \quad \Longrightarrow \quad \delta \mathcal{N}_{\mu\nu} = 0,
\]
where $\mathcal{N}_{\mu\nu} = 0$ is the Neumann condition, and $\delta\mathcal{N}_{\mu\nu}$ the linearized version.
Thus, fluctuations keep us in a ``code subspace'' \cite{Almheiri2014} of branes obeying Neumann conditions.\footnote{Note that this condition can also trivially be satisfied by setting all variations at the brane to zero, i.e., by choosing Dirichlet boundary conditions. We will however focus on the dynamical case.} 
Moreover, we expect the vanishing of the $\chi_A$ integral to hold for more general bulk Lagrangians, and therefore to give a simple means to determine consistent linearized equations of motion for a brane immersed in a bulk gravity theory with higher-derivative terms or scalar fields.

The outline of this note is as follows.  In section 2 we review background material, including BCFTs and the proposed bottom-up holographic dual, along with the first law of entanglement entropy and its relation to the bulk Einstein equations. In section 3 we introduce the covariant phase space formalism, which is our main technical tool. In section 4 we compute the bulk equations of motion in the presence of a boundary. We end with discussion and directions for future research.

\subsection*{Notation}

We will use $\ext$ for spacetime exterior derivatives and $\delta$ for configuration space exterior derivatives. The interior product between a spacetime vector $\xi$ and spacetime form $\omega$ will be denoted by $\xi \cdot \omega$, while the interior product between a configuration space vector $\Xi$ and a configuration space form $\omega$ is given by $\iota_\Xi \omega$. Generally, spacetime vector fields are denoted by lowercase Greek letters and configuration space vector fields by uppercase Greek letters.

\vspace{10pt}
\section{Preliminaries}
We start by briefly outlining some useful background material.
\vspace{10pt}
\subsection{AdS/BCFT}
\label{sec:adsbcft}

We consider a $d$-dimensional CFT on a flat half-space $\mathbf{H}_d
:= \{x\in \mathbb{R}^{1, d-1}:x^1 \geq 0\}$, equipped with a
conformally invariant boundary condition $B$. 
In Lorentzian signature, this breaks the global symmetry group from $\text{SO}(d,2)$ to $\text{SO}(d-1,2)$ \cite{McAvity1995}.
Since $\text{SO}(d-1,2)$ is the isometry group of $\text{AdS}_d$, the
natural semiclassical dual $M_{d+1}$ is the \emph{Janus metric}, where we foliate the bulk with warped copies of $\text{AdS}_d$ \cite{Bak2003, Gutperle2012}:
\begin{align}
\label{janus1}
\begin{split}
  \ext s_M^2 & = f^2(\mu) \left(\ext\mu^2 + \ext s_{\AdS_{d}}^2\right)  \\
   & = f^2(\mu) \left[ \ext\mu^2 + \frac{-\ext t^2 + \ext r_{||}^2 + r_{||}^2 \, \ext\Omega_{d-3}^2+ \ext z^2}{z^2} \right] \\
    &=f^2(\mu) \left[ \ext\mu^2 + \frac{-\ext t^2 + \ext\rho^2 + \rho^2
      \sin^2 \phi \, \ext\Omega_{d-3}^2+ \rho^2\, \ext\phi^2}{\rho^2
      \cos^2 \phi} \right].
\end{split}
\end{align}
Here, $r_{||}$ is the radial coordinate on the defect.
Slices are parameterized by $\mu \in [0,\pi]$, with $\mathbf{H}_d$ at
$\mu = 0$.
We have also introduced polar coordinates $(\rho, \phi)$ for $\AdS_d$,
with $ (r_{||}, z) =: \rho(\sin\phi, \cos\phi)$.
The natural $(d+1)$-dimensional holographic coordinate $Z$ 
and other coordinates obey the relations\footnote{This
  follows by choosing the conventional defining function $C(Z) =
  Z^2/L_\AdS^2$ and placing the BCFT on a flat background, $\ext
  s_{\text{BCFT}}^2 = -\ext t^2 + \ext (x^1)^2 + \ext r_{||}^2 +
  r_{||}^2\, \ext \Omega_{d-3}^2$.
  For further discussion, see \cite{Jensen2013}.
}

\begin{equation}
  \label{coord-rels}
Z := z \sin \mu, \quad x^1 = z\cos \mu \quad \Longrightarrow \quad r^2
= \rho^2 (\sin^2 \phi + \cos^2\phi \cos^2\mu).
\end{equation}
Pure $\AdS_{d+1}$ has warp factor $f_\text{AdS}(\mu) :=
L_\AdS\sin^{-1}(\mu)$, so the metric has denominator $Z^2$ \cite{Gutperle2012}.
To recover the usual AdS/CFT correspondence far from the boundary, the
warp factor $f$ must approach $f_\text{AdS}$ as $\mu \to 0$.
For the purposes of this work we will set $L_\AdS = 1$.
Departures from $f_\text{AdS}(\mu)$ at $\mu > 0$ correspond to stress-energy in the bulk.
These can arise, even in the vacuum state of the BCFT, when the
boundary condition switches on $\text{SO}(d-1,2)$-invariant sources
for bulk fields.

The Janus slicing follows from the symmetry of the BCFT vacuum.
The new ingredient in the ETW prescription for AdS/BCFT is a codimension-1 hypersurface $\mathcal{B}_d$ which terminates the bulk geometry.
To maintain $\text{SO}(d-1,2)$ symmetry, the brane must be a
particular $\text{AdS}_d$ slice located at $\mu = \mu_B$.
We emphasize that, for a localized brane, this is the only choice
consistent with symmetry.
Our warping parameter is then restricted to $\mu \in [0, \mu_B]$ in (\ref{janus1}).
It is clear from the metric that $\mathcal{B}_d$ will be a hypersurface of constant extrinsic curvature:
\begin{equation}
  \label{eq:4}
  K(\mu_B) := h_{ab}\partial_n h^{ab}|_{\mu=\mu_B} = \frac{f'(\mu_B)}{f(\mu_B)}
\end{equation}
where $h_{ab} = g_{ab}|_{\mu=\mu_B}$ is the induced metric on the slice, $a, b$ are coordinates tangential to the brane, and $\partial_n$ is the normal derivative.

\begin{figure}[t]
 \centering
 \includegraphics[scale=0.37]{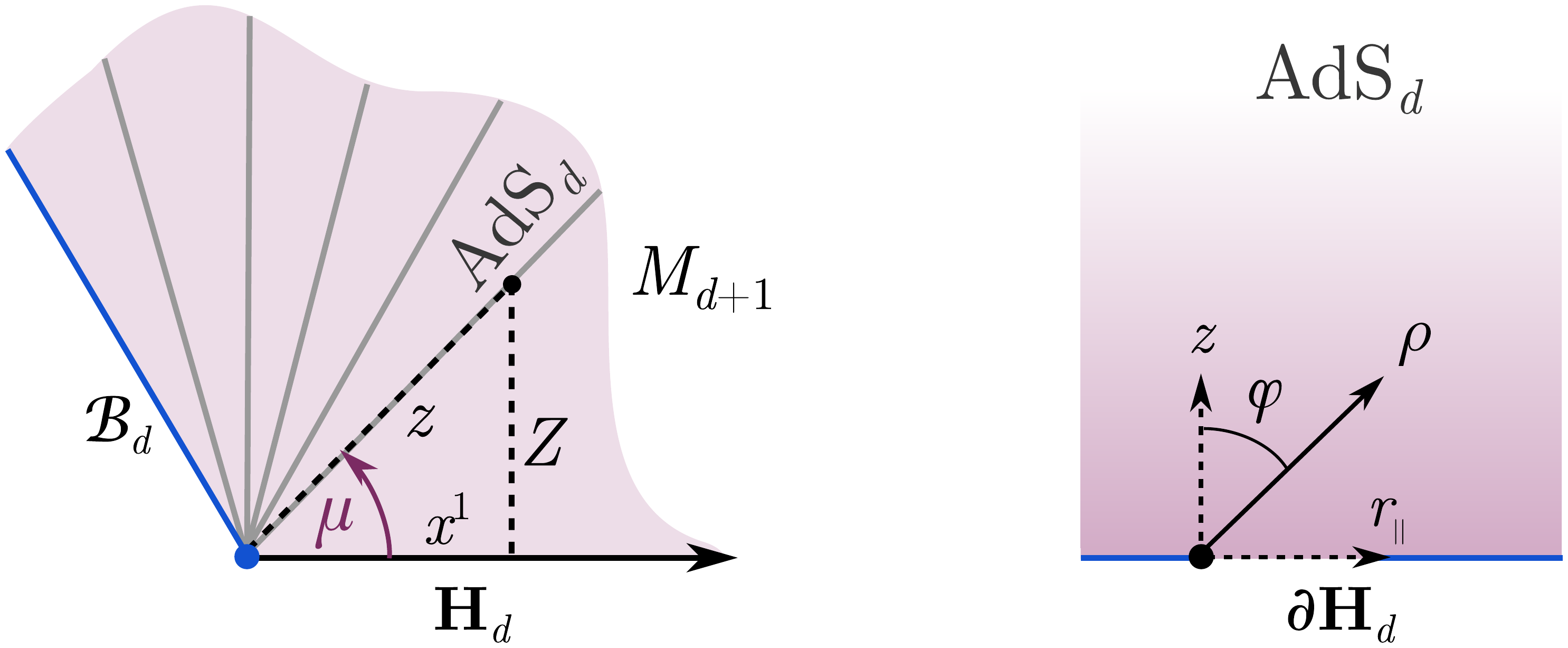}
\vspace{-10pt}
  \caption{\emph{Left}. Janus slicing for bulk dual $M_{d+1}$ of a $\text{BCFT}_d$. The warp factor is represented by a purple envelope. \emph{Right}. Coordinates for $\text{AdS}_d$, with time and $\Omega_{d-3}$ directions suppressed. 
}
\label{fig:coords}
\end{figure}

We can force the brane to sit at a location of constant extrinsic curvature by adding an additional term
to the action.
Usually, the \emph{Gibbons-Hawking-York} (GHY) boundary term,
\begin{equation}
  \label{ghy}
  I_\text{GHY} = -\frac{1}{8\pi G^{(d+1)}_\text{N}} \int \ext^{d-1}y\, \sqrt{|h|}K,
\end{equation}
is evaluated on a fixed spatial boundary and used to regulate the bulk action
This corresponds to \emph{Dirichlet conditions} where we fix $h$ and let the bulk solution determine the embedding, hence $K$.
Alternatively, we can consider \emph{Neumann conditions} where $h$ is dynamical.
Extremizing $I_\text{GHY}$ determines its equation of motion.
For $d \neq 1$, this gives $K_{\mu\nu} = h_{\mu\nu} K$ which implies
$K = 0$.
We can easily modify the GHY term to obtain nonzero extrinsic curvature by adding a \emph{tension }term on the brane:
\begin{equation}
  \label{eq:7}
  I_\text{GHY} = -\frac{1}{8\pi G^{(d+1)}_\text{N}} \int \ext^{d-1}y \sqrt{|h|}(K + T).
\end{equation}
The equations of motion for $d \neq 1$ then become \cite{Cooper2018}
\begin{equation}
  \label{brane-eom}
  K_{\mu\nu} - h_{\mu\nu}(K+T) = 0 \quad \Longrightarrow \quad  K = \frac{d}{(1-d)}T.
\end{equation}
The tension is essentially a cosmological constant on the brane.
For general matter on the brane, $Th_{\mu\nu}$ is replaced by a
brane-localized stress-energy $T_{\mu\nu}^{\text{brane}}$, but for the purposes of this paper, we
focus on the constant tension case.

\vspace{10pt}
\subsection{The first law of entanglement}
\label{sec:first-law-entangl}

We now turn to entanglement measures.
Consider a Hilbert space $\mathcal{H}$ and some density matrix $\rho$
on $\mathcal{H}$.
Each density matrix is associated with a \emph{modular Hamiltonian} $H_\rho$
and \emph{von Neumann entropy} $S(\rho)$, defined by $\rho =:
e^{H_\rho}/\tr[e^{H_\rho}]$ and $S(\rho) := -\mbox{tr}(\rho \log \rho)$.
In a quantum field theory, the von Neumann entropy for a spatial
subregion generally diverges due to short-distance effects.
If we define a reference state $\sigma$ on $\mathcal{H}$, a better-behaved measure is the \emph{relative entropy}
\begin{equation}
  \label{rel-ent}
  S(\rho || \sigma) := \tr(\rho \log \rho) - \tr(\rho \log \sigma),
\end{equation}
which is finite since the UV divergences cancel.
Relative entropy has the useful property of being positive-definite
\cite{Witten2018}, with $S(\rho \| \sigma) \geq 0$ and $S(\rho \|
\sigma) = 0$ just in case $\rho = \sigma$.
Note that we can rewrite (\ref{rel-ent}) as a difference in von
Neumann entropies and expectations of vacuum modular Hamiltonians:
\begin{align}
S(\rho || \sigma) & = \big[\tr(\rho \log \rho) - \tr(\sigma \log \sigma)\big] + \big[\tr(\sigma \log \sigma) - \tr(\rho \log \sigma)\big] \notag \\
  & = S(\sigma) - S(\rho) + \langle H_\sigma \rangle_\rho - \langle H_\sigma \rangle_\sigma = -\Delta S + \Delta \langle H_\sigma\rangle,
\end{align}
where $\Delta X := X(\rho)- X(\sigma)$ for any function $X$.

Define $\rho := \sigma + \delta\rho$ as a small perturbation of our reference state.
Positive-definiteness implies that the relative entropy is at least quadratic in $\delta \rho$, $S(\rho||\sigma) = O\big(\delta \rho^2\big)$, and hence to leading order
\begin{align}
  \label{first-law}
  \delta S = \delta \langle H_\sigma\rangle.
\end{align}
This is the \emph{first law of entanglement}.
It states that to linear order in the perturbation $\delta \rho$, the
change in von Neumann entropy equals the change in the expectation value of the modular Hamiltonian defined with respect to $\sigma$.

We now specialize to the vacuum state $|B\rangle \in \mathcal{H}_B$ of
a BCFT.\footnote{We assume that the backreaction is entirely captured
  by the warping function $f(\mu)$.}
For any spatial subregion $A$ of the BCFT, we can factorize the Hilbert space into degrees of freedom inside and outside $A$, $\mathcal{H} = \mathcal{H}_A \otimes \mathcal{H}_{\bar{A}}$.\footnote{In fact, short-distance divergences make such a factorization impossible in field theory. However, it is a convenient fiction, and yields the same results as the more circuitous but correct route of factorizing states via Tomita-Takesaki theory \cite{Witten2018}.}
Let $\mathbb{B}^{+}_a(R)$ denote a half-ball of radius $R$ centered at
some boundary point $a \in \partial \mathbf{H}_d$, i.e., bisected by
the boundary.
We will take $a = 0$ for simplicity.
Define the reference state as the reduced density matrix on this ball:
\[
\sigma := \mbox{tr}_{\bar{A}} |B\rangle\langle B|.
\]
Our first task is to describe the modular Hamiltonian.

A boundary-centered half-ball $\mathbb{B}_0^+(R)$ is related by $\mathbb{Z}_2$ symmetry to a full ball, and most results carry over from the usual CFT case immediately.
In particular, the modular Hamiltonian is \cite{Casini2018}
\begin{align}
  \label{mod-ham}
  H_\sigma = \int_{\Sigma^+} \ext^{d-1}x\, \eta^\mu \zeta^\nu T_{\mu\nu},
\end{align}
where $\eta^\mu$ is a timelike unit vector normal to
$\mathbb{B}^+_{0}(R)$, and $\zeta^\nu$ is the conformal Killing vector associated
with the conformal transformation keeping $\partial\mathbb{B}^+_0(R)$ fixed:
\begin{align}
  \label{zeta}
  \zeta(t, x^i) = \frac{\pi}{R}\left[\big(R^2 - t^2-r^2\big)\partial_t
  -2 t r \partial_r\right].
\end{align}
This is proved using the same conformal map to a thermal state on a
hyperbolic CFT as Casini-Huerta-Myers used to prove the Ryu-Takayanagi
(RT) formula \cite{Ryu2006a} for ball-shaped regions
\cite{Casini2011}.
The boundary of the BCFT maps to a uniformly accelerated surface \cite{Jensen2013}.

\vspace{10pt}
\subsection{Holographic entanglement entropy}
\label{sec:hologr-entangl-entr}

The thermality of the reduced state in hyperbolic coordinates can be
turned into a proof of the RT formula \cite{Jensen2013}.
The boundary Killing field (\ref{zeta}) corresponds to a \emph{bulk} timelike Killing vector
\begin{equation}
  \xi(t, x^i, \rho) = \frac{\pi}{R}\left[\big(R^2 - t^2 - \rho^2\big)\partial_t - 2
    t \rho \partial_\rho \right],\label{xi}
\end{equation}
for any warping profile $f(\mu)$.
This approaches $\zeta$ as $\mu \to 0$.
In turn, this is associated with a hyperbolic black hole in the bulk, whose
Killing horizon lies at $\xi = 0$, or equivalently $\rho = R$.
The thermal entropy of the BCFT state is computed by the
Bekenstein-Hawking entropy $\mathcal{A}/4G_\text{N}$ of the
Killing horizon.

It is straightforward to show that this Killing horizon is also the
\emph{minimal area} surface approaching the half-ball region at the
boundary \cite{Jensen2013}.
This proves the RT formula for half-balls in a BCFT:
\begin{equation}
  S(\sigma) = \frac{\mathcal{A}_\text{min}}{4G_\text{N}} .\label{rt-bcft}
\end{equation}
The usual homology condition for the RT formula states that the
boundary region $A$ and bulk surface $\mathcal{S}_A$ together bound some
bulk spacelike codimension-1 region $\Xi_A$:
\begin{equation}
\partial \Xi_A = A \cup \mathcal{S}_A.\label{hom-1}
\end{equation}
For half-balls in a BCFT dual to AdS with an ETW brane, the RT surface $\rho = R$ indicates that this condition is modified:
\begin{equation}
  \label{hom-2}
  \partial \Xi_A = A \cup \mathcal{S}_A \cup \mathcal{B}_A,
\end{equation}
where $\mathcal{B}_A$ is some portion of the ETW brane $\mathcal{B}_d$ between $\mathcal{S}_A$ and $A$. As first pointed out by Headrick \cite{Headrick2013}, this implies that the minimal surface is normal to the brane. This is the prescription adopted in \cite{Cooper2018}, albeit for time-dependent configurations.

We note the close resemblance between this apparent
violation of the homology condition and the ``quantum extremal
island'' proposal of AMMZ for evaporating black holes
\cite{Almheiri2019TheHole}.
In that context, a 2D BCFT is coupled to a quantum system dual to a
2D black hole.
This black hole acts as a ``Planck brane'' in the UV, similar to the
ETW brane in the dual of the BCFT.
The AMMZ proposal allows for candidate extremal surfaces to end on
some ``extremal island'' $\mathcal{I}_g$ of the brane, just as the RT surface ends
on some region $\mathcal{B}_A$ of the ETW brane in our case.
For a related discussion, where the evaporating
black hole is modelled with a BCFT, see \cite{Rozali:2019day}.

This is the story for Einstein gravity.
For a covariant Lagrangian $\mathcal{L}$
describing some other theory of gravity (with higher-curvature
corrections for instance), the black hole
entropy is given by evaluating the \emph{Wald functional} on the horizon \cite{Wald1993, Iyer1994}:
\begin{equation}
  \label{wald-func}
  S^\text{W} = -2\pi \int_{\mathcal{S}} \ext^{d-1} \sigma \,
  \sqrt{h} \frac{\delta \mathcal{L}}{\delta R^{ab}{}_{cd}} n^{ab}n_{cd},
\end{equation}
where $h$ is the induced metric, and $n_{ab}:= n^{(1)}_an^{(2)}_b -
n^{(1)}_bn^{(2)}_a$ is the horizon binormal, built out of unit vectors
$n^{(1,2)}$ orthogonal to each other and the horizon.
Recall that the \emph{surface gravity} $\kappa$ of a black hole, with Killing
horizon generated by $\xi$, is defined by
\begin{equation}
  \label{surface-gravity}
  \nabla_{[c}\xi_{d]} = \kappa  n_{cd}.
\end{equation}
The Killing vector $\xi$ in (\ref{xi}) is normalized such that $\kappa
= 2\pi$.

\begin{figure}[t]
 \centering
 \includegraphics[scale=0.4]{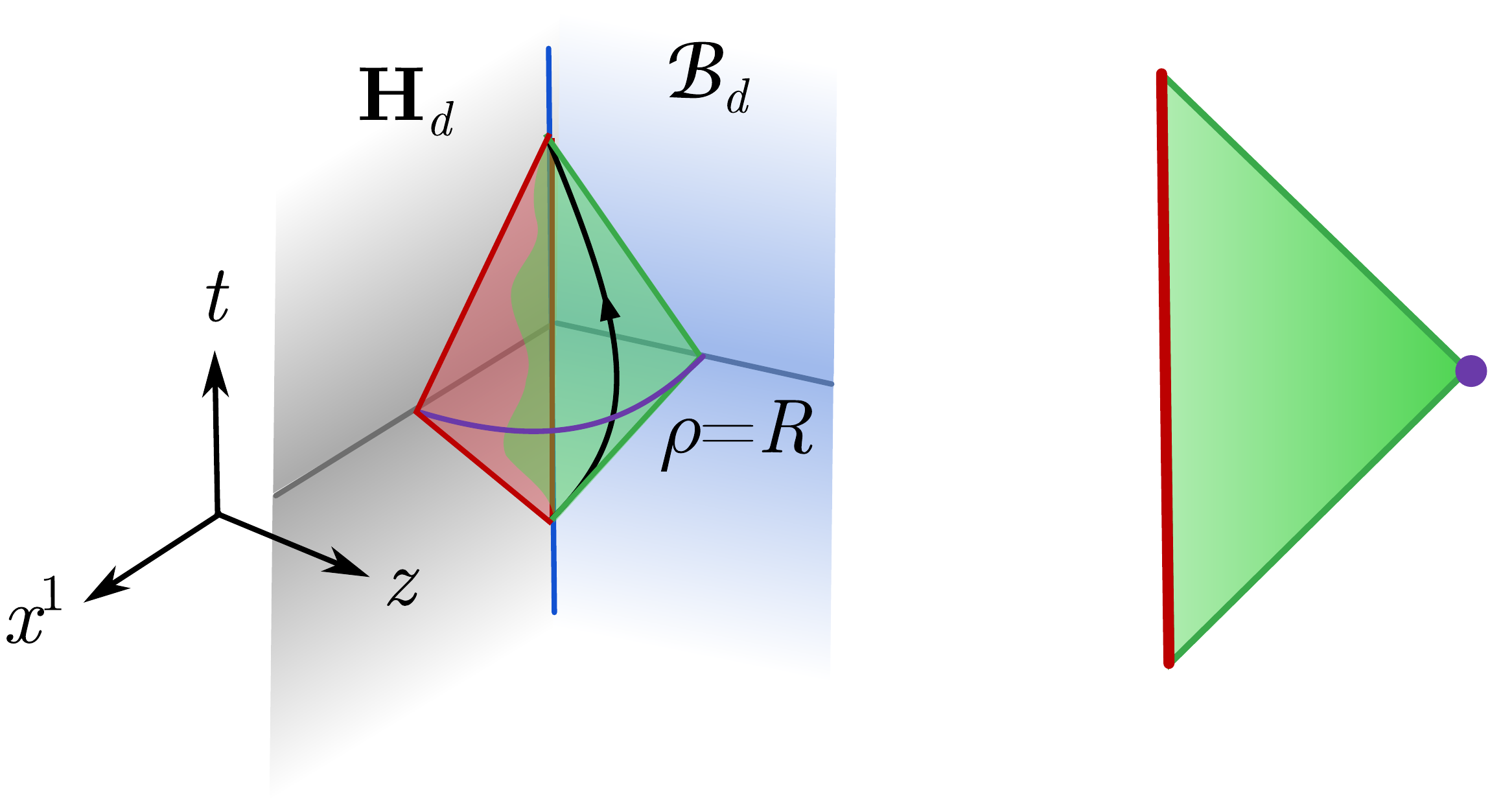}
\vspace{-10pt}
  \caption{\emph{Left}. The bulk causal domain
    $\tilde{\mathcal{D}}[\mathbb{B}^+]$ (green), cut away to reveal
    the boundary causal domain $\mathcal{D}[\mathbb{B}^+]$ (red). The
    minimal surface is purple. \emph{Right}. The Penrose diagram
    for the hyperbolic black hole. The minimal surface and bifurcation
    surface (purple) coincide.
}
\label{fig:bulk}
\end{figure}

Consider a perturbation to the state of our theory which corresponds to a perturbation of the metric in the bulk, $g \to g + \delta g$.
To first order in $\delta g$, the change in the entanglement entropy for a half-ball
$\mathbb{B}^+$ is given by the change in (\ref{wald-func}):
\begin{equation}
\delta S_{\mathbb{B}^+} = \delta S_{\mathbb{B}^+}^\text{W},\label{delta-S}
\end{equation}
where $S^\text{W}$ is the Wald functional evaluated on the
\emph{unperturbed} minimal surface $\mathcal{S}_{\mathbb{B}^+}$.\footnote{Evaluating on the perturbed surface produces second-order
corrections. Moreover, it is known that the Wald
functional does not produce the entanglement entropy of an arbitrary
boundary region in higher-derivative gravity due to differences in the
universal terms \cite{Dong:2013qoa}.
These corrections are quadratic in the extrinsic curvature of
$\mathcal{S}_A$, and hence vanish for the Killing horizon associated
with the half-ball $A = \mathbb{B}^+$.}
The modular Hamiltonian (\ref{mod-ham}) involves an integral over the
boundary stress-energy, so under a change of state dual to the change
of metric,
\begin{equation}
\delta \langle H_{\mathbb{B}^+}\rangle = \int_{\Sigma^+} \ext
\Sigma^\mu
\zeta^\nu \delta\langle T_{\mu\nu}\rangle.\label{delta-H}
\end{equation}
Near the boundary, and away from the brane, we can use the usual
Fefferman-Graham coordinate system with coordinate $z$, and expand $\delta g_{ab} := z^{d-2} h^{(d)}_{ab} + O(z^{d-1})$.
The variation in CFT stress-tensor expectation is proportional to this
leading piece projected onto the boundary, $\delta\langle T_{\mu\nu}\rangle = C h^{(d)}_{\mu\nu}$, and hence
\begin{equation}
  \label{fefferman-graham}
  \delta \langle H_{\mathbb{B}^+}\rangle = C\int_{\Sigma^+} \ext \Sigma^\mu \zeta^\nu h^{(d)}_{\mu\nu}.
\end{equation}
This is well-known for Einstein gravity, but holds more generally \cite{Faulkner2014}.
Even with the identifications (\ref{delta-S}--\ref{fefferman-graham}), the first
law (\ref{first-law}) need not be satisfied for arbitrary $\delta
g$.
The main result of \cite{Faulkner2014} is that the first law
(\ref{first-law}) for ball-shaped regions $\mathbb{B}$ in the CFT implies that
perturbations around the AdS vacuum obey linearized equations of
motion. Perhaps this is unsurprising when we have defined both sides holographically in terms of the Wald entropy. But the energy variation only knows about the metric near the boundary, while the entropy variation knows about the deep bulk. We require a condition on $\delta g$ to ensure these two variations agree.

Suppose there exists a $(d-1)$-form $\chi_{\mathbb{B}}$ with the properties that
\begin{equation}
  \label{chi}
  \int_{\mathbb{B}} \chi_{\mathbb{B}} = \delta \langle H_{\mathbb{B}}\rangle, \quad \int_{\mathcal{S}_\mathbb{B}} \chi_{\mathbb{B}} = \delta S_{\mathbb{B}},
\end{equation}
where $\mathcal{S}_\mathbb{B}$ is the extremal surface associated with
$\mathbb{B}$.
In addition, suppose the (spacetime) exterior derivative of $\chi$
vanishes when $\delta g$ is on-shell, i.e., $\ext \chi \propto \delta
E$, where $\delta E$ are the linearized equations of motion.
Then for $\Xi_{\mathbb{B}}$ in
(\ref{hom-1}), the first law follows from Stokes theorem:
\begin{equation}
  0 = \int_{\Xi_{\mathbb{B}}} \ext \chi = \int_{\mathbb{B} -
    \mathcal{S}_{\mathbb{B}}} \chi = \delta S_\mathbb{B} - \delta
  \langle H_\mathbb{B}\rangle.\label{stokes-first-law}
\end{equation}
We defer the definition and detailed treatment of $\chi_{\mathbb{B}}$ to the next section.

\vspace{0pt}
\begin{figure}[t]
 \centering
 \includegraphics[scale=0.45]{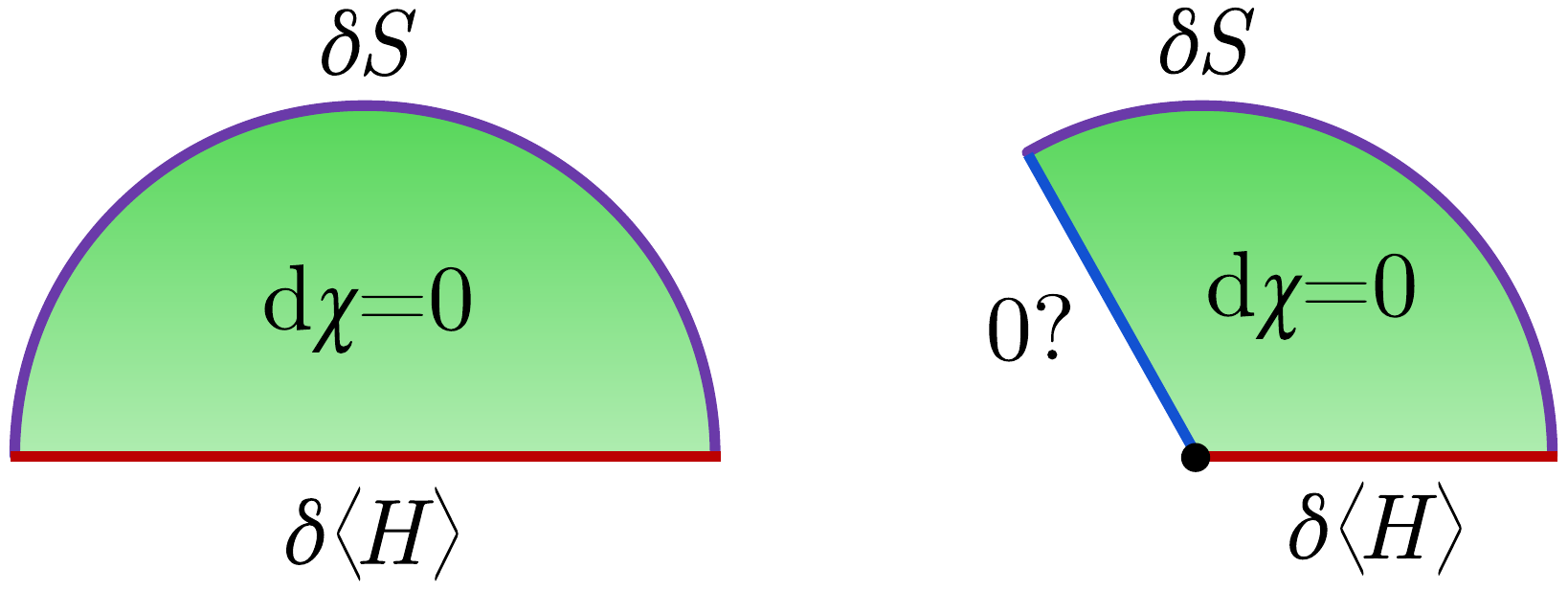}
\vspace{-10pt}
  \caption{\emph{Left}. The first law from Stokes theorem in
    AdS/CFT. \emph{Right}. For AdS/BCFT, the first law requires that the contribution from the brane vanishes.
}
\end{figure}

In the BCFT, the homology condition is modified to (\ref{hom-2}).
Even if we can construct a $(d-1)$-form which is exact on-shell, the
integral (\ref{stokes-first-law}) will become
\begin{equation}
\int_{\mathcal{B}_{\mathbb{B}^+}} \chi_{\mathbb{B}^+} = \delta S_{\mathbb{B}^+} - \delta \langle
  H_{\mathbb{B}^+}\rangle.\label{brane-first-law}
\end{equation}
Since the energy (calculated from the half-ball modular Hamiltonian)
and entropy (calculated from the bulk black hole horizon) are fixed,
the first law \emph{requires} that the integral over the brane
vanishes.
We will see that this enforces a Neumann condition in Einstein gravity. 
This resembles the logic of gravitation from entanglement in CFTs: the
first law places a constraint on integrated metric fluctuations, which
in turn is equivalent to local linearized dynamics.

\vspace{10pt}
\section{Covariant phase space formalism}
According to Noether's theorem, every continuous symmetry yields a conserved current, with an associated charge generating the symmetry transformation via the Poisson bracket (classical mechanics) or commutator (quantum mechanics). 
In the Hamiltonian formalism, defining these brackets breaks spacetime covariance by selecting a preferred time-slicing. 
For diffeomorphism invariant theories, 
the \emph{covariant phase space formalism} \cite{Crnkovic:1987tz, Lee1990,Iyer1994,Wald1999, Harlow2019}
provides an alternative.
This endows the space of solutions with a natural symplectic form $\Omega$ \cite{Crnkovic:1987tz}, whose inverse $\Omega^{-1}$ is the Poisson bracket in classical theories, and the commutator in quantum theories.

Let $\mathcal{P}$ be our phase space of solutions, which is a subset of all possible field configurations. 
Suppose that $\mathcal{P}$
is also equipped with a symplectic form: a closed, nondegenerate 2-form $\Omega \in \Lambda^2(\mathcal{P})$.
Closure means $\delta \Omega = 0$, where $\delta$ is the exterior derivative in the space of field configurations, while the nondegeneracy condition is
\begin{equation}
 \Omega(X, Y) = 0 \text{ for all $Y$} \Longrightarrow X = 0.
\end{equation}
This induces a map from tangent vectors on $\mathcal{P}$ to one-forms:
\begin{equation}
  \label{eq:5}
  X \mapsto \omega_X := \Omega(\cdot, X) = -\iota_X \Omega,
\end{equation}
where the minus sign arises from anticommuting $X$ into the second slot.
By nondegeneracy, this map is invertible, with $\Omega^{-1}(\cdot, \omega_X) = X$.

Consider a continuous symmetry along a flow $\xi$ in spacetime, generated by a charge $H_\xi:\mathcal{P}\to\mathbb{R}$.
A familiar example is time translation and the Hamiltonian $H$.
To define the action on phase space, we first construct the (phase space) vector field dual to $\delta H_\xi$:
\begin{align}
\Xi := \Omega^{-1}(\cdot, \delta H_\xi), \quad \delta H_\xi = \Omega(\cdot, \Xi).
\label{Xi}
\end{align}
For any function $f: \mathcal{P}\to\mathbb{R}$, the infinitesimal variation is then
\begin{align}
\delta_\xi f := \iota_\Xi \delta f = \Omega^{-1}(\delta f, \delta H_\xi).
\label{poiss}
\end{align}
This is Hamilton's equation $\dot{f} = \{f, H\}$ in covariant phase space language.
We should caution the reader that $\delta f$ is a one-form on configuration space, while $\delta_\xi f$ maps points in phase space (solutions to the equations of motion) to functions on spacetime (variations of $f$).
For a general configuration-space differential form $\omega$, we define the variation under $\xi$ using the Lie derivative:
\begin{equation}
  \label{field-var}
  \delta_\xi \omega := \mathcal{L}_\Xi \omega.
\end{equation}
This agrees with (\ref{poiss}) for a 0-form $f$.
Our goal in this section is to find an expression for the generator of infinitesimal transformations, $\delta H_\xi$, in the presence of a boundary.

\vspace{10pt}
\subsection{The symplectic form}
\begin{figure}[t]
 \centering
 \includegraphics[scale=0.35]{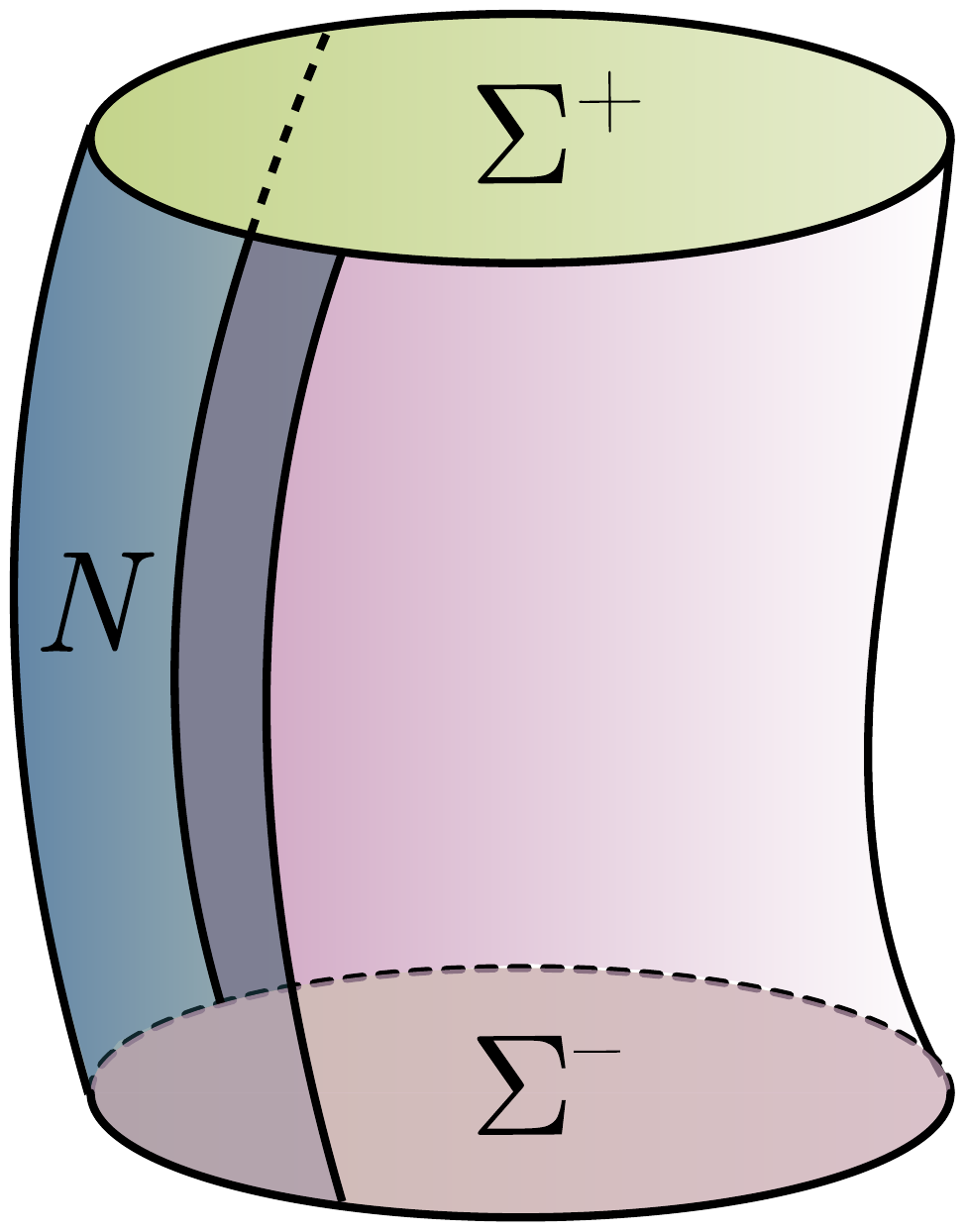}
\vspace{-0pt}
  \caption{
A manifold $M$ with boundary $\partial M = \Sigma^+ \cup \Sigma^- \cup \Gamma$, and timelike region $N \subset \Gamma$ coupling to bulk fields.
}
\label{fig:symp}
\end{figure}

The boundary of our manifold $M_{d+1}$ has a region asymptotic to the BCFT as well as an ETW brane, with $\partial M_{d+1} = \mathbf{H}_d \cup \mathcal{B}_d$.
Even in the vacuum boundary state $|B\rangle$, bulk fields can be switched on, which can be holographically modelled with bulk field sources on the brane \cite{Fujita2011}.
We therefore consider the more general scenario of a manifold $M$ with boundary
\begin{equation}
\partial M_{d+1} = \Sigma^+ \cup \Sigma^- \cup \Gamma, \quad N \subset \Gamma,\label{eq:6}
\end{equation}
where $\Sigma^\pm$ are past a future spacelike boundaries, $\Gamma$ is the timelike boundary, and $N$ is the region with bulk couplings, depicted in Figure \ref{fig:symp}.

In addition to the bulk action, and boundary terms required for a well-defined variational principle, the coupling on $N$ will require a new term in the action:
\begin{align}
    \label{eq:action_with_boundary_coupling}
    S := \int_M L + \int_{\partial M} \ell +  \int_N \ell'.
\end{align}
From this definition, we see that $L$ is a top-level $(d+1)$-form in spacetime, $\ell, \ell'$ are spacetime $d$-forms, and in configuration space, $L, \ell, \ell'$ are simply functions.
For the purposes of this paper, it will be sufficient to assume that $\ell'$ depends only on the bulk fields evaluated on $N$ and the normal vector $n^\mu$ at the boundary.
In particular, we assume that the boundary Lagrangian $\ell'$ does not depend on derivatives of the bulk fields, or on additional degrees of freedom localized to $N$.\footnote{We leave this extension to future work.}
This means we can write $\delta \ell' =: t(\phi) \, \delta \phi$, where $\phi$ stands for bulk fields (including the metric).

To build a symplectic form, we first build a phase space from solutions to the equations of motion.
These are defined as stationary points of the action, with field configurations on the past and future boundaries $\Sigma^\pm$ fixed.
This means that only terms local to $\Sigma^\pm$ contribute to the variation.
An infinitesimal variation of the action gives
\begin{align}
    \delta S = - \int_M E(\phi) \,\delta \phi + \int_{\Gamma} (\Theta + \delta \ell) + \int_{N} \delta \ell' + \int_{\Sigma^+ - \Sigma^-} (\Theta + \delta \ell),
\label{gen-field-var}
\end{align}
where $\Theta$ is the boundary term arising from varying $L$:
\begin{equation}
  \label{vary-L}
  \delta L =: -E(\phi) \, \delta \phi + \ext\Theta.
\end{equation}
Stationarity of the action requires the timelike contributions from $\Gamma$ and $N$ cancel.
Let us seek boundary conditions which ensure this.

Our action $S$ should be functionally differentiable away from $\Sigma^\pm$ \cite{Regge1974}.
From $\delta \ell' = -t(\phi)\, \delta\phi$, we can massage the timelike integrals into the form
\begin{align}
    \int_{\Gamma} (\Theta + \delta \ell) + \int_{N} \delta \ell' = \int_{\Gamma} \big(\ext C- e(\phi) \,\delta \phi\big) - \int_{N} t(\phi) \,\delta \phi,
\end{align}
where $e(\phi)$ is implicitly defined by this equation.
The $\ext C$ is easily dealt with, since we can evaluate it on $\Sigma^\pm$ after a change of sign.\footnote{Note that by Stokes' theorem, we can also evaluate $C$ on $\partial \Gamma = \partial \Sigma^\pm$. This is manifestly independent of of how we extend $\ell$ into the bulk.}
To force the $\delta \phi$ terms to vanish, however, we must choose appropriate boundary conditions.
For a boundary $\Gamma$ at finite distance, one option is \emph{Dirichlet conditions} $\delta\phi|_\Gamma = 0$.
Alternatively, we can allow for near-boundary dynamics by imposing $e + t = 0$ (on $N$) or $e = 0$ (on $\Gamma\backslash N$) as equations of motion.
If part of the boundary is asymptotic, we require these equations to vanish sufficiently quickly as we approach infinity.

Suppose we have chosen boundary conditions ensuring the $\Gamma\backslash N$ term vanishes, for instance, the canonical choice of Dirichlet conditions on the AdS boundary, while not making any statment for the situation on the ETW brane.
In this case, the general variation of $S$ takes the form
\begin{align}
    \delta S = - \int_M E(\phi) \,\delta \phi - \int_N \big( e(\phi) + t(\phi)\big) \delta\phi + \int_{\Sigma^+ - \Sigma^-} (\Theta + \delta \ell - \ext C).
\label{general-var-S}
\end{align}
The \emph{pre-symplectic potential} $\omega$ is the exterior derivative of the last integrand:
\begin{align}
    \omega := \delta (\Theta + \delta \ell - \ext C) = \delta \left( \Theta - \ext C \right),
\label{pre-symb-pot}
\end{align}
since $\delta^2 \ell = 0$.
The \emph{pre-symplectic form} $\tilde \Omega$ is the integral of $\omega$ over some Cauchy slice $\Sigma$:
\begin{equation}
  \tilde{\Omega} := \int_\Sigma \omega.
\end{equation}
In order to obtain phase space proper, we must quotient out the zero modes, defining $\mathcal{P} := \tilde{\mathcal{P}}/\tilde{\mathcal{G}}$, on which $\tilde{\Omega}$ lifts to a genuine symplectic form $\Omega$ \cite{Harlow2019}.

\vspace{10pt}
\subsection{The generator of infinitesimal diffeomorphisms}
We now consider theories which are invariant under arbitrary diffeomorphisms preserving the boundary conditions.
The symmetry acts on configuration space according to (\ref{poiss}) or (\ref{field-var}).
In order for the theory to be covariant, the spacetime variation under a diffeomorphism $\xi$ should \emph{induce} the corresponding field variation:
\begin{align}
   \delta_\xi L = \mathcal L_\xi L.
\end{align}
In general, a configuration space form $\omega$ transforms covariantly if the action on the fields, induced by the flow $\Xi$, agrees with the full spacetime variation under $\xi$:
\begin{equation}
\delta_\xi \omega = \mathcal L_\xi \omega.\label{eq:12}
\end{equation}
By a theorem of Iyer and Wald \cite{Iyer1994}, if $L$ is covariant, then the integration by parts used to define $\Theta$ can be performed covariantly, and hence $\Theta$ can be taken to be covariant.

By Cartan's magic formula
\begin{equation}
\mathcal{L}_X \omega = X \cdot \ext \omega + \ext (X \cdot \omega),\label{cartan}
\end{equation}
the Lie derivative of $L$ reduces to an exterior derivative
\begin{align}
\mathcal{L}_\xi L =  \xi \cdot \ext L + \ext (\xi \cdot L) = \ext (\xi \cdot L),
\label{ext-L}
\end{align}
where $\ext L = 0$ since $L$ is a top-level spacetime form.
Similar statements hold for the $d$-forms $\ell, \ell'$ on $\partial M$.
The action (\ref{eq:action_with_boundary_coupling}) transforms as
\begin{align}
    \delta_\xi S &= \int_M \mathcal L_\xi L + \int_{\partial M} \mathcal L_\xi \ell + \int_N \mathcal L_\xi \ell' = \int_{\partial M} \xi \cdot L + \int_{\partial N} \xi \cdot \ell',
\end{align}
using Stokes' theorem and $\partial^2 M_{d+1} =0$.
Equating this to an insertion $\iota_\Xi \delta S$ in the general field variation (\ref{gen-field-var}), we find that for a diffeomorphism-covariant theory,
\begin{align}
\int_M E(\phi) \,\delta_\xi \phi & = \int_N \delta_\xi \ell' + \int_{\partial M} \delta_\xi \ell+ \int_{\partial M} \iota_\Xi\Theta - \int_{\partial M} \xi \cdot L - \int_{\partial N} \xi \cdot \ell'
\notag \\ & = \int_{\partial M} (\iota_\Xi \Theta - \xi \cdot L).
\label{eq:off-shell_current}
\end{align}
The contributions from $\ell'$ cancel, while the $\ell$ integral vanishes.
If the equations of motion hold, the left hand side of (\ref{eq:off-shell_current}) is zero, suggesting we define the \emph{Noether current} 
\begin{align}
    J_\xi := \iota_\Xi \Theta - \xi \cdot L.
\end{align}
A quick calculation using (\ref{vary-L}), (\ref{ext-L}), and $\ext \delta = \delta \ext$ shows that, for a covariant Lagrangian,
\[
\ext J_\xi = -\mathcal{L}_\xi L + \delta_\xi L   + E(\phi)\delta_\xi \phi =  E(\phi)\delta_\xi \phi,
\]
so $J_\xi$ is conserved on-shell.
When the result is true for arbitrary diffeomorphisms $\xi$, the on-shell conservation $\ext J_\xi = 0$ implies \cite{Wald1990} the existence of a $(d-2)$-form Noether ``charge'' $Q_\xi$, defined by
\begin{equation}
  \label{dQ}
  J_\xi = \ext Q_\xi.
\end{equation}
The derivation of $J_\xi$ depends only on bulk equations of motion, and is insensitive to the tension of the brane. 
The ``charge'' $Q_\xi$ arises due to \emph{bulk} diffeomorphism invariance, which is also independent of boundary conditions.

We can now derive the infinitesimal generator $\delta H_\xi$.
First, we take the exterior derivative of $J_\xi$ in configuration space, using (\ref{pre-symb-pot}), Cartan's formula (\ref{cartan}) for both $\delta$ and $\ext$, (\ref{vary-L}), the covariance of $\Theta$, and the fact that $\delta$ commutes with spacetime insertions:
\begin{align}
\delta J_\xi & = -\xi \cdot \delta L + \delta(\iota_\Xi \Theta) \notag \\
& = \xi \cdot (E(\phi) \delta \phi - \ext \Theta) - \iota_\Xi \delta \Theta + \mathcal{L}_\Xi \Theta \notag\\
& = \xi \cdot E(\phi) \delta \phi - \mathcal{L}_\xi \Theta + \mathcal{L}_\Xi \Theta + \ext (\xi \cdot \Theta) - \iota_\Xi \delta \Theta \notag\\
& = \xi \cdot E(\phi) \delta \phi + \ext (\xi \cdot \Theta) - \iota_\Xi \omega - \iota_\Xi \delta \ext C.
\end{align}
Integrating over a Cauchy slice $\Sigma$, we find that
\begin{align}
    \delta \int_\Sigma J_\xi + \int_{\partial \Sigma} (\iota_\Xi \delta C - \xi \cdot \Theta )= -\iota_\Xi \Omega + \int_\Sigma \xi \cdot E(\phi) \delta \phi.
\label{charge-symp}
\end{align}
For a diffeomorphism-invariant theory, and a family of solutions with $E(\phi) = 0$, (\ref{dQ}) and the definition (\ref{Xi}) lets us extract $\delta H_\xi$ from (\ref{charge-symp}):
\begin{align}
    \label{eq:generator_boundary_integral}
    \delta H_\xi = -\iota_\Xi\Omega = \int_{\partial\Sigma} (\delta Q_\xi - \xi \cdot \Theta + \iota_\Xi \delta C). 
\end{align}
Thus, we have an expression for the generator of infinitesimal diffeomorphisms as a boundary integral.
The result is independent of the boundary coupling Lagrangian $\ell'$, reflecting the insensitivity of the Noether current to boundary tension.

\subsection{The first law revisited}
\label{sec:killing}

If $\xi$ is a Killing vector field, then the solution should be \emph{invariant} under $\xi$, and the associated vector field $\Xi$ on phase space vanishes.
From (\ref{eq:generator_boundary_integral}), it follows that
\begin{align}
    \label{eq:vanishing_generator}
    \int_{\partial \Sigma} (\delta Q_\xi - \xi \cdot \Theta + \iota_\Xi \delta C) = 0.
\end{align}
Although $\iota_\Xi \delta C = 0$ for a Killing symmetry, we will keep
the term for reasons of mathematical nicety in the next section.
In the context of AdS/BCFT, the modular flow on the causal domain of a half-ball is associated with a Killing field (\ref{xi}).
But recall from \S \ref{sec:hologr-entangl-entr} that the first law instructs us to find a form $\chi$ which is (spacetime) exact on-shell, and as per (\ref{chi}), yields the modular energy and entropy when evaluated on the boundary and extremal surface respectively.
The form
\begin{equation}
  \chi_\xi := \delta Q_\xi - \xi \cdot \Theta
\end{equation}
is indeed exact on-shell, since
\begin{align}
  \ext \chi_\xi & = \delta \ext Q_\xi - \ext (\xi \cdot \Theta) = \delta J_\xi - \ext (\xi \cdot \Theta) = 0.
\end{align}
It remains to check the boundary and horizon limit of $\chi_\xi$.
The reasoning is similar to the CFT case \cite{Faulkner2014}, but we briefly outline the argument for completeness.

Let us first consider the Killing horizon associated to the half-ball, $\mathcal{S}_{\mathbb{B}^+}$.
At the horizon, the field $\xi$ vanishes by definition, so $\chi_\xi|_{\mathcal{S}_{\mathbb{B}^+}} = \delta Q_\xi$.

To connect this to the entropy, we will exploit a technical result due to Iyer and Wald \cite{Iyer1994}.
If $\xi$ is normalized so that $\kappa = 2\pi$, they proved that the Wald entropy (\ref{wald-func}) equals the integral of the Noether ``charge'':
\begin{equation}
  \label{noether-wald}
  S^{\text{W}}_{\mathbb{B}^+} = \int_{\mathcal{S}_{\mathbb{B}^+}} Q_\xi.
\end{equation}
Then, using (\ref{delta-S}) and the normalisation of $\xi$, we have, as required,
\begin{equation}
  \delta S_{\mathbb{B}^+} = \delta S_{\mathbb{B}^+}^\text{W}  = \int_{\mathcal{S}_{\mathbb{B}^+}} \delta Q_\xi = \int_{\mathcal{S}_{\mathbb{B}^+}}\chi_\xi.
\end{equation}
The horizon integral gives the variation in entropy.

We now consider the boundary integral over the half-ball
$\mathbb{B}^+$.
It can be shown \cite{Faulkner2014} that this integral reduces to (\ref{fefferman-graham}), and hence
\begin{equation}
  \delta \langle H_{\mathbb{B}^+}\rangle = \int_{\mathbb{B}^+} \chi_\xi = \int_{\mathbb{B}^+} \big(\delta Q_\xi - \zeta \cdot \Theta\big),
\end{equation}
using $\xi \to \zeta$ at the boundary.
It follows that, when linearized perturbations are on-shell
in the bulk with $E(\phi) = 0$, we have
\[
\delta S[\xi] - \delta \langle H_{\mathbb{B}^+}\rangle = \int_{\mathcal{B}_{\mathbb{B}^+}} \chi_\xi.
\]
Thus, the first law $\delta S = \delta \langle H\rangle$ implies that
the brane contribution vanishes:
\begin{equation}
\int_{\mathcal{B}_{\mathbb{B}^+}} \chi_\xi = 0.\label{brane-vanish}
\end{equation}
This statement holds for the broad class of covariant bulk Lagrangians
and bulk couplings we are considering.

In the preceding argument, we have implicitly assumed that there is no additional delta-localized stress-energy source (or more precisely, variation thereof) on the boundary of the CFT.
Since we consider linearized metric fluctuations only, and do not introduce any boundary-localized fields, this seems like a reasonable assumption.

Furthermore, as the authors of \cite{Herzog2017b} have argued, one should expect that in a quantum theory, the non-conservation of a boundary stress-energy tensor ``thickens'' boundary degrees of freedom, such that the theory contains only one single stress energy tensor without delta-function support on the boundary. This operator is conserved and the normal-transverse component to the brane, $T^{na}|_{\mathcal{B}} = 0$ vanishes. The non-trivial normal-normal component of the stress-energy tensor becomes the \emph{displacement operator} \cite{Herzog2017a} as we approach the boundary. However, this component does not appear in the present discussion. 

It might still be that the expectation value of the CFT stress-energy tensor contains delta-function terms localized to the boundary. In fact, generally such terms appear in the Weyl anomaly of a BCFT. However, since we assume a flat boundary such terms to not contribute and by energy-momentum conservation, localized stress-energy at the boundary does not require any other treatment than localized stress-energy at any other point.

In light of recent progress on the black hole information paradox \cite{Almheiri2019a}, it would interesting to understand how this relates to toy models where quantum mechanical degrees of freedom dual to gravity are localized on the boundary of the CFT. We leave this to future work.

\vspace{10pt}
\section{Equations of motion from the first law}
The integrals along the extremal surface and boundary, discussed in the previous section, give the variation in entanglement entropy and modular energy appearing on either side of the first law.
Restricting to metric perturbations which satisfy the first law implies that the brane contribution (\ref{brane-vanish}) vanishes. In principle, this can be used to reverse engineer linearized equations of motion for the brane in any covariant theory of gravity. We will consider the familiar case of Einstein
gravity as a proof of concept.

\subsection{Generator at the ETW brane}
In coordinates, the first two terms of (\ref{eq:vanishing_generator}) read
\begin{align}
\label{eq:closed_integral}
\begin{split}  
\left( \delta Q_\xi - \xi \cdot \Theta \right)_{\alpha \beta\cdots} = \frac{\epsilon_{\alpha \beta \cdots\gamma \delta}}{16 \pi G_N} \Big( & \delta g^{\gamma \lambda} \nabla_\lambda \xi^\delta - \frac 1 2 \delta g_{\;\;\lambda}^\lambda \nabla^\gamma \xi^\delta - \xi^\lambda \nabla^\gamma \delta g^\delta_{\;\;\lambda} \\ &   - g^{\gamma \lambda} \xi^\delta \nabla^\kappa \delta g_{\lambda \kappa} + g^{\gamma \lambda} \xi^\delta \nabla_\lambda \delta g^\kappa_{\;\;\kappa} \Big),
\end{split}
\end{align}
where the $\cdots$ stand for additional transverse directions.
Since we are integrating over a codimension-2 region, we can choose a two-dimensional basis of vectors normal to the path of integration, with $\gamma$ the index parallel to the brane and $\delta$ normal to it. We will calculate the expression piece by piece.

\begin{itemize}
\item \textbf{Parallel terms.}
First, we consider terms with index $\gamma$. Using this, along with the fact that at the brane, $\xi^\mu n_\mu = 0$ and $h^{\mu\nu} \xi_\nu = \xi^\mu$, we find
\begin{align}
    \label{eq:first_contribution_raw}
    \begin{split}
    \left( \delta Q_\xi - \xi \cdot \Theta \right)_{\alpha \beta\cdots} \overset{(1)}{\supset} \frac{\epsilon_{\alpha \beta \cdots\gamma \delta} }{16 \pi G_N} n^\delta h^{\gamma \mu} \xi^\nu \Big(& - \delta g_{\;\;\mu}^{\lambda} K_{\lambda\nu} + \delta g_{\mu\lambda} n^\lambda n^\alpha n^\beta \nabla_\alpha \xi_\beta \\ &+ \frac 1 2 \delta g_{\;\;\lambda}^\lambda K_{\mu \nu} - n^\lambda \nabla_\mu \delta g_{\nu \lambda} \Big),
    \end{split}
\end{align}
where $K_{\mu\nu}$ is the extrinsic curvature.
To streamline the notation, we will use a SVT--like decomposition of
the metric fluctuations:
\begin{align}
\gamma_{\mu\nu} & := h_\mu^{\;\;\alpha} \delta g_{\alpha \beta}
h^\beta_{\;\;\nu} \\ A_\mu &: = h_\mu^{\;\;\alpha} \delta g_{\alpha
  \beta} n^\beta \\ \phi & := n^{\alpha} \delta g_{\alpha \beta}
n^\beta.
\end{align}
Reorganising all of (\ref{eq:first_contribution_raw}) into these terms, we arrive at
\begin{align}
    \label{eq:first_contribution_SVT}
    \begin{split}
    \left( \delta Q_\xi - \xi \cdot \Theta \right)_{\alpha \beta\cdots} \overset{(1)}{\supset} \frac{\epsilon_{\alpha \beta \cdots\gamma \delta} }{16 \pi G_N} n^\delta h^{\gamma \mu} \xi^\nu \Bigg[ - \gamma_{\;\;\mu}^\alpha K_{\alpha \nu} + \frac 1 2 ( \phi + \gamma^\lambda_{\;\;\lambda}) K_{\mu \nu} - n^\lambda \nabla_\mu \delta g_{\nu \lambda} \Bigg] \\
    + \frac{\epsilon_{\alpha \beta \cdots\gamma \delta} }{16 \pi G_N} n^\delta h^{\gamma \mu} \Big( A_\mu n^\alpha n^\beta \nabla_\alpha \xi_\beta \Big).
    \end{split}
\end{align}

\item \textbf{Normal terms.}
We next consider terms which arise from exchanging $\gamma$ and $\delta$ in (\ref{eq:closed_integral}). We can write this as
\begin{align}
\label{eq:second_contribution_raw}
\begin{split}
     \left( \delta Q_\xi - \xi \cdot \Theta \right)_{\alpha \beta\cdots} \overset{(2)}{\supset} {}& \frac{\epsilon_{\alpha \beta \cdots\gamma \delta} }{16 \pi G_N} n^\delta h^{\gamma\mu} \xi_\nu \Bigg[ h_{\mu\nu}  n^\rho h^{\alpha\beta }  ( \nabla_{\alpha} \delta g_{\rho \beta} - \nabla_{\rho} g_{\alpha\beta} ) + n^\rho \nabla_\rho \delta g_{\mu \nu} \Bigg]  \\ & \qquad + \frac{\epsilon_{\alpha \beta \cdots\gamma \delta} }{16 \pi G_N} n^\delta h^{\gamma\mu} \Bigg[ \frac 1 2 (\gamma^\lambda_{\;\;\lambda} - \phi) n^\rho \nabla_\rho \xi_\mu - A^a \mathcal D_a \xi_\mu \Bigg],
\end{split}
\end{align}
and therefore find
\begin{align}
\begin{split}
    \label{eq:second_contribution_SVT}
    \left( \delta Q_\xi - \xi \cdot \Theta \right)_{\alpha \beta\cdots} \overset{(2)}{\supset} \frac{\epsilon_{\alpha \beta \cdots\gamma \delta} }{16 \pi G_N} n^\delta h^{\gamma\mu} \xi^{\nu}\Bigg[ h_{\mu\nu} h^{\alpha\beta} n^\rho ( \nabla_\alpha \delta g_{\rho \beta} + \nabla_\beta \delta g_{\rho \alpha}- \nabla_\rho \delta g_{\alpha\beta} )  \\+ n^\rho \nabla_\rho \delta g_{\mu \nu} + \frac 1 2 \phi K_{\mu\nu} - \frac 1 2 \gamma_{\;\;\lambda}^\lambda  K_{\mu\nu} + h_{\mu\nu} K^{\alpha\beta} \gamma_{\alpha\beta} - h_{\mu\nu} K \phi\Bigg] \\  - n^\delta h^{\gamma\mu} \Bigg[\frac 1 2 \phi n^\rho \partial_\rho \xi_\mu - \frac 1 2 \gamma^\lambda_{\;\;\lambda} n^\rho \partial_\rho \xi_\mu + \mathcal D_a(  A^a \xi_\mu) \Bigg].
    \end{split}
\end{align}
Above, $\mathcal D$ denotes the covariant derivative on the brane, and
we have used that
\[
h_{\alpha_1}^{\;\;\alpha_2} \dots h_{\alpha_{n-1}}^{\;\; \alpha_{n}}
  \nabla_{\alpha_2} T_{\alpha_4 \dots \alpha_n} =
  h_{\alpha_1}^{\;\;\alpha_2} \dots h_{\alpha_{n-1}}^{\;\; \alpha_{n}}
  \mathcal D_{\alpha_2} T_{\alpha_4 \dots \alpha_n}
\]
 if $T$ is transverse to the brane.

\item \textbf{A convenient zero.}
Finally, we consider $ \iota_\Xi \delta C$.
This vanishes for a Killing field $\xi$.
However, if we evaluate for an \emph{arbitrary} $\xi$, we obtain $0$
in a convenient form that simplifies the contributions above.

\hspace{15pt}Recall that $\delta C$ is a spacetime $(d-1)$-form and a phase-space one-form. In our coordinates, and again using our convention that $\gamma$ is the coordinate direction parallel to the brane but orthogonal to the direction of integration, it reads \cite{Harlow2019}
\begin{align}
    \delta C = \delta_2 (\sqrt{-h} h^{\gamma \nu} \delta_1 g_{\nu\mu} n^\mu),
\end{align}
where we have to antisymmetrize $\delta_1$ and $\delta_2$.
In more detail:
\begin{align}
    \begin{split}
    \delta_2 (\sqrt{-h} h^{\gamma \nu} \delta_1 g_{\nu\mu} n^\mu) =
     \sqrt{-h} h^{\gamma \nu} \Bigg[ \frac 1 2 h^{\alpha\beta} \delta_2 g_{\alpha\beta} \delta_1 g_{\nu\mu} n^\mu  - \delta_2 g_{\nu\mu} h^{\mu\alpha} \delta_1 g_{\alpha\beta} n^\beta  \\ - \,\,h^{\mu\alpha} \delta_2 g_{\alpha\beta} n^\beta \delta_1 g_{\nu\mu}  - \frac 1 2 n^\alpha \delta_2 g_{\alpha\beta} n^\beta \delta_1 g_{\nu\mu} n^\mu  \Bigg] - (1 \leftrightarrow 2).
     \end{split}
\end{align}
Due to this antisymmetrization, the second and third terms cancel. 
Antisymmetrizing, calculating the interior product $\iota_\Xi
\delta C$, and replacing $\delta_2 g_{\mu\nu}$ by $\mathcal L_\xi
g_{\mu\nu}$, after some algebra we obtain
\begin{align}
\label{eq:third_contribution_SVT}
    \iota_\Xi\delta C = \sqrt{-h} h^{\gamma \mu} \Bigg[ A_\mu (\mathcal D_\alpha \xi^\alpha -
  n^\alpha n^\beta \nabla_\alpha \xi_\beta ) + \frac 1 2
  (\gamma^\alpha_{\;\; \alpha  } - \phi) (\xi^\nu K_{\nu\mu} - n^\nu
  \nabla_\nu \xi_\mu) \Bigg].
\end{align}
\item \textbf{The final expression.}
To obtain the final expression, we add
(\ref{eq:first_contribution_SVT}), (\ref{eq:second_contribution_SVT}),
and (\ref{eq:third_contribution_SVT}):
\begin{align}
    \begin{split}
(\chi_\xi)_{\alpha\beta\cdots}
=  \frac{\epsilon_{\alpha \beta \cdots\gamma \delta} }{16 \pi G_N} n^\delta h^{\gamma\mu} \xi^\nu \Big[ h_{\mu\nu}  n^\rho (h^{\alpha\beta} - h^{\alpha}_{\mu}h^{\beta}_{\nu})  ( \nabla_{\alpha} \delta g_{\rho \beta} - \nabla_{\rho} \delta g_{\alpha\beta} )  \\  - \gamma_{\;\;\mu}^\alpha K_{\alpha \nu} + \gamma^\lambda_{\;\;\lambda} K_{\mu \nu} \Big] \\ 
       \qquad + \frac{\epsilon_{\alpha \beta \cdots\gamma \delta} }{16 \pi G_N} n^\delta h^{\gamma\mu} \big(  - A^a \mathcal D_a \xi_\mu \big) +
           \sqrt{-h} h^{c \mu} \big( A_\mu \mathcal D_\alpha \xi^\alpha  \big).
    \end{split}
\label{final-exp}
\end{align}
\item \textbf{The Neumann condition.}
\noindent This seems unwieldy, but is directly related to the Neumann condition on the brane.
Using the identity
\begin{align}
    0 = - h_\mu^\alpha h_\nu^\beta n^\rho \nabla_\beta \delta g_{\rho \alpha} + \mathcal D_\nu A_\mu + \phi K_{\mu\nu} - K_\nu^\lambda \gamma_{\lambda \mu},
\end{align}
we can rewrite (\ref{final-exp}) as
\begin{align}
\label{almost}
\begin{split}
(\chi_\xi)_{\alpha\beta\cdots}= \frac{\epsilon_{\alpha \beta \cdots\gamma \delta} }{16 \pi G_N} n^\delta h^{\gamma\mu} \xi^{\nu}\Big[ n^\rho (h_{\mu\nu} h^{\alpha\beta} - h_\mu^\alpha h_\nu^\beta)  ( \nabla_\alpha \delta g_{\rho \beta} + \nabla_\beta \delta g_{\rho \alpha}- \nabla_\rho \delta g_{\alpha\beta} ) \\ + \phi (K_{\mu\nu} - h_{\mu\nu} K) + h_{\mu\nu} K^{\alpha\beta} \gamma_{\alpha\beta} + K_{\mu\nu} \gamma^\lambda_{\;\;\lambda} - 2 \gamma^\kappa _\mu K_{\kappa\nu} \Big] \\  - n^\delta h^{\gamma\mu} \mathcal D_\kappa(  A^\kappa \xi_\mu - \xi^\kappa A_\mu) .
\end{split}
\end{align}
If the background obeys the brane equation of motion
(\ref{brane-eom}), it follows that
\begin{align}
    h_{\mu\nu} K^{\alpha\beta} \gamma_{\alpha\beta}  + \gamma_{\alpha\beta} h^{\alpha\beta}K_{\nu\mu} - 2 \gamma^\kappa _\mu K_{\kappa\nu} = 2 \big[h_{\mu\nu} K^{\alpha\beta} - h_\mu^\alpha h_\nu^\beta (K+T)\big] \gamma_{\alpha\beta}.
\end{align}
This turns equation (\ref{almost}) into an
expression
proportional to a linearized Neumann condition, plus some additional terms depending on $A_\mu$.
Luckily, we can eliminate these by choosing normal coordinates close
to the brane.
In these coordinates, we have
  \begin{align}
    \begin{split}
      \label{eq:final_expression_SVT}
(\chi_\xi)_{\alpha\beta\cdots} =
\frac{\epsilon_{\alpha \beta \cdots\gamma \delta}}{16\pi G_N}n^\delta
h^{\gamma\mu} \xi^{\nu}\delta \mathcal{N}_{\mu\nu},
    \end{split}
  \end{align}
where the Neumann condition at first order (derived in Appendix \ref{app:neumann_bc_at_first_order}) is: 
  \begin{align}
  \begin{split}
    0 = \delta \mathcal{N}_{\mu\nu} := n^\rho (h_{\mu\nu} h^{\alpha\beta} - h_\mu^\alpha h_\nu^\beta)
    ( \nabla_\alpha \delta g_{\rho \beta} + \nabla_\beta \delta
    g_{\rho \alpha}- \nabla_\rho \delta g_{\alpha\beta} ) \\ +
    (K_{\mu\nu} - h_{\mu\nu} K) \phi + 2 \big[h_{\mu\nu} K^{\alpha\beta} -
    h_\mu^\alpha h_\nu^\beta (K+T) \big] \gamma_{\alpha\beta}.
  \end{split}
\label{neumann-first}
  \end{align}
\end{itemize}

\vspace{10pt}
\subsection{Local equations of motion}
Combining the explicit expression (\ref{eq:final_expression_SVT}), and the first law in the form (\ref{brane-vanish}), we learn that
\be
\int_{\mathcal{B}_A} \chi_A \propto \int_{\mathcal{B}_A} h^{\gamma \mu} \xi^\nu \delta \mathcal{N}_{\mu \nu} n^\delta= 0 \label{eq:branevanish}
\ee
for any boundary-centered half-ball $A$.
We will see that this set of global conditions implies the local condition $\delta \mathcal{N}_{\mu\nu} = 0$ everywhere on the brane.
This is analogous to \cite{Faulkner2014}, where the first law gives global constraints equivalent to local, linearized bulk equations of motion.

\begin{figure}[t]
\centering
\subfigure{\label{fig:cauchyNtt}\includegraphics[width=77mm]{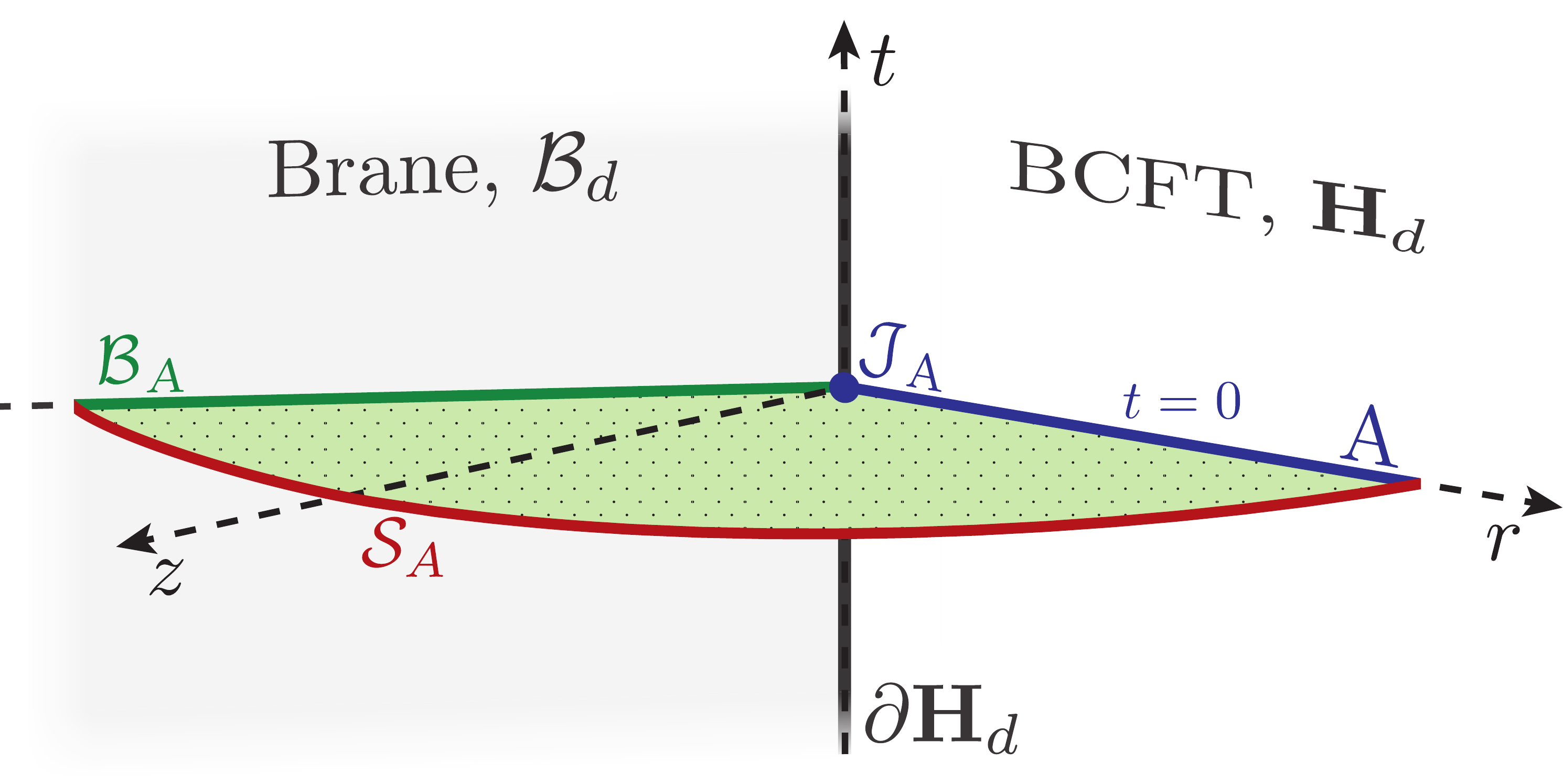}}
\subfigure{\label{fig:cauchyNtz}\includegraphics[width=77mm]{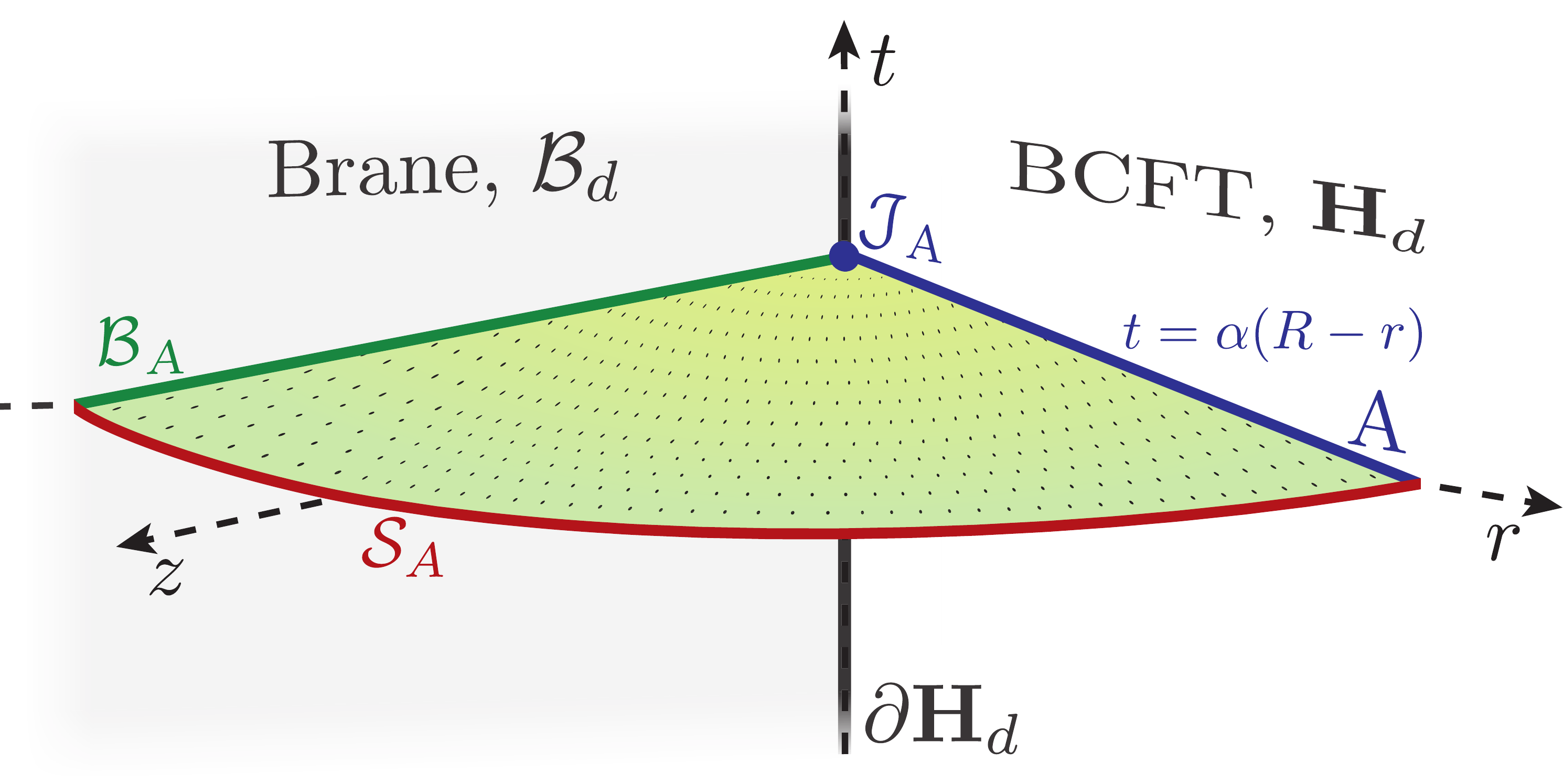}}
\caption{\emph{Left:} The Cauchy slice, with $t=0$, used to show that the integral of $\delta \mathcal{N}_{tt}$ on $\mathcal{B_A}$ vanishes.\newline
\emph{Right:} The Cauchy slice, with $t=\alpha(R-r)$, used to show that the integral of $\delta \mathcal{N}_{t\rho}$ on $\mathcal{B_A}$ vanishes.}
\label{fig:cauchyNs}
\end{figure}

First, we can use the trick in \cite{Faulkner2014} to trade global constraints on $\chi_A$, which depends on boundary region $A$, for global constraints on $\delta\mathcal{N}_{\mu\nu}$.
Choosing our Cauchy slice to intersect $t=0$ on the brane (Figure \ref{fig:cauchyNs}), and applying the differential operator $R^{-1} \partial_R R$ to (\ref{eq:branevanish}), produces two terms: an integral of $\chi_A$ localised to $\rho=R$, where the Killing vector $\xi_A$ is zero by definition, and another term involving the derivative of the Killing vector.
Noting that both $h^{\gamma\mu}$ and $\xi_A^\nu$ project onto the timelike direction, we find
\be
\int_{\mathcal{B}_A} \delta \mathcal N_{t t} = 0.\label{eq:calcNtt}
\ee
For the standard $t = 0$ Cauchy slice, the regions $\mathcal{B}_A$ are hemisphere-shaped, and we can invoke the argument from Appendix A of \cite{Faulkner2014} to conclude that\footnote{One way to see this is by taking derivatives of (\ref{eq:calcNtt}) with respect to the variables parametrizing $\mathcal{B}_A$, namely its radius $R$ and center $x_0^i$.
Applying $\partial_R$ or $\partial_{x_0^i}$ respectively gives
\be
\int_{\partial \mathcal{S}_A \cap \partial \mathcal{B}_A} \delta \mathcal N_{t t} = \int_{\partial \mathcal{S}_A \cap \partial \mathcal{B}_A} x^i \delta \mathcal N_{t t} = 0.\label{eq:diffNtt}
\ee
We can repeat this process, substituting the integrands of (\ref{eq:diffNtt}) into (\ref{eq:calcNtt}), 
to conclude that the overlap integral of $\delta \mathcal N_{t t}$ with an arbitrary polynomial in $x^i$ vanishes.
Since $\delta \mathcal N_{t t}$ is continuous, it can be approximated to arbitrary precision by such a polynomial, and the vanishing overlaps imply $\delta\mathcal{N}_{tt}=0$ on $\partial \mathcal{S}_A \cap \partial \mathcal{B}_A$, since the integral of $(\delta\mathcal{N}_{tt})^2$ vanishes to arbitrary precision.
These semi-circles $\partial \mathcal{S}_A \cap \partial \mathcal{B}_A$ cover the brane, so we must have $\delta \mathcal N_{t t} = 0$ everywhere.}
\be
\delta \mathcal N_{t t} = 0.
\ee

By boosting along the BCFT boundary, we then find that $\delta \mathcal{N}_{u u} = u^\mu u^\nu \delta \mathcal{N}_{\mu \nu} = 0$ vanishes for any timelike vector $u$ along this boundary, implying that $\delta \mathcal{N}_{ab} = 0$ for all $a, b$ parallel to this boundary. We still need to show that the components of $\delta \mathcal{N}$ in the $z$-direction (or equivalently $\rho$) vanish. This is achieved by choosing our Cauchy slices to lie at $t = \alpha(R-r)$, with $0 \leq \alpha < 1$. In our case, this is in fact possible by conservation of the stress-energy tensor, even at the boundary. Using this slicing, and applying $\partial_R R$ to (\ref{eq:branevanish}), we arrive at\footnote{Here we omit and additional factor of $1/\sqrt{1-\alpha^2}$ that can be divided out.}
\be
\int_{\mathcal{B}_A} \{ [2 R(1-\alpha^2) + 2 r \alpha^2] \partial_t - 2 \alpha \rho \partial_\rho \}^\mu \delta \mathcal{N}_{\mu \nu} \{ \partial_r + \alpha \partial_\rho \}^\nu = 0.\label{eq:calcNtz}
\ee
After taking the limit $\alpha\to0$, and applying another derivative with respect to $R$, we find
\be
\int_{\mathcal{B}_A} \delta \mathcal{N}_{t \rho} = 0.
\ee
By the same argument as above, $\delta \mathcal{N}_{t \rho} = 0$ everywhere on the brane. Since the other components vanish, we immediately have $\delta \mathcal{N}_{t z} = 0$. 
Substituting into (\ref{eq:calcNtz}), we deduce that $\delta \mathcal{N}_{\rho \rho} \propto \delta \mathcal{N}_{z z}$ vanishes everywhere on the brane.
Assembling the components, we find the local equation of motion $\delta\mathcal{N}_{\mu\nu} = 0$ everywhere on the brane. If we exclude the trivial case of Dirichlet conditions, which set all metric fluctuations close to the brane to zero, we conclude that metric fluctuations must obey Neumann boundary conditions.

\vspace{10pt}
\section{Discussion}

In this note, we have probed the consequences of the first law of entanglement for boundary CFTs dual to AdS cut off by an ETW brane.
We find that, for a broad class of gravitational Lagrangians, the first law implies that bulk Noether charges associated with boundary-centered half-balls must vanish when integrated over associated regions of the brane. For the specific case of Einstein gravity in the bulk, and perturbations to a brane of constant curvature, this vanishing implies local, linearized equations of motion $\delta \mathcal{N}_{\mu\nu}=0$, where $\mathcal{N}_{\mu\nu} = 0$ is the background Neumann condition. 
In other words: if the brane obeys a background Neumann condition, the first law implies that non-vanishing metric fluctuations obey a linearized Neumann condition at the brane. 

Our main tool is that for half-ball shaped regions in the BCFT, both the entanglement entropy and the vacuum modular Hamiltonian are known.
This closely parallels the logic of \cite{Faulkner2014}, where a background obeying Einstein's equations (or some more general covariant equations of motion) are forced to obey linearized equations by the first law. Our method should generalize to give local, linearized equations for perturbations to the brane in more general theories of gravity.

We have focused on pure gravity, where the only relevant feature of the boundary state is its tension, sourcing the identity sector of the BCFT. 
Although warping partially captures the effect of (unperturbed) background fields, it would be interesting to perturb light fields in the bulk and determine
the restrictions arising from the corresponding statement of the first law \cite{deBoer2016}. We expect to find that these light bulk fields obey linearized equations sourced by the ETW brane. Using these methods, it may even be possible to map a consistent boundary state, satisfying some explicit holographic restrictions, to a solution of bulk equations in the presence of a localized source. This would be the equivalent to \cite{Heemskerk:2009pn} for a BCFT.

Similarly, one might envision additional fields or dynamical gravity on the brane itself.
This presumably requires an even more careful treatment of the boundary contributions to the Noether charge and the effects of additional degrees of freedom localized to the BCFT boundary. While it did not seem to play an important role in our discussion, a Hayward term might need to be included \cite{Hayward:1993my} (for recent work, see e.g. \cite{Takayanagi:2019tvn}). A thorough analysis could lead to an extended holographic dictionary between bulk modes and fields localized on the brane. 

This note has shown that quantum information-theoretic considerations, successfully applied in AdS/CFT to constrain bulk dynamics \cite{Faulkner2014} and semiclassical states \cite{Lashkari:2016idm}, can also be used to constrain the dynamics of holographic BCFTs.
It would be interesting to consider holographic constraints on excited states of BCFTs, e.g., positive energy theorems in the bulk \cite{Lashkari:2016idm, Neuenfeld2018} , but we leave this question, and the sundry extensions mentioned above, to future work.

\vspace{10pt}
\section*{Acknowledgements}
We are thankful to Jamie Sully and Mark Van Raamsdonk for useful discussions. MR and SC are supported by a Discovery grant from NSERC.
DN is partially supported by the University of British Columbia through a Four Year Fellowship and the Simons Foundation through an It from Qubit Postdoctoral Fellowship. 
DW is supported by an International Doctoral Fellowship from the University of British Columbia.
Research at Perimeter Institute is supported in part by the Government of Canada through the Department of Innovation, Science and Economic Development Canada and by the Province of Ontario through the Ministry of Economic Development, Job Creation and Trade.

\vspace{10pt}
\appendix 

\vspace{10pt}
\section{Useful formulae and conventions}
\label{app:formulae}
We can decompose the metric close to the brane into two parts:
\begin{align}
    g_{\mu\nu} = n_{\mu} n_\nu + h_{\mu\nu},
\end{align}
where $n_\mu$ is the normal vector to the boundary with $n \cdot n =
1$ and $h_{\mu\nu}$ is orthogonal to $n_\mu$, $n^\mu h_{\mu\nu} = 0$.
Since the covariant derivative is metric compatible, we have that $\nabla_\alpha (n^\mu n^\nu + h^{\mu\nu}) = 0$. Contracting this equation with $n_\mu$ and $h_{\beta\nu}$ and also projecting $\alpha$ onto the transverse space we obtain the useful identity
\begin{align}
    K_{\alpha\beta} = - h_{\beta \nu} n_\mu \nabla_\alpha h^{\mu\nu}.
\end{align}
Finally, we need some standard results for the variation of $h$ and
$n$ under perturbations to $g$:
\begin{align}
    \sqrt{-h} = \frac 1 2 \sqrt{-h} h^{\alpha\beta} \delta g_{\alpha \beta} \\
    \delta (h^{c \nu}) = - h^{c \alpha} \delta g_{\alpha\beta} h^{\beta \nu} \\
    \delta(n^\mu) = -g^{\mu\rho} \delta g_{\rho\nu} n^\nu + \frac 1 2 n^\mu n^\alpha \delta g_{\alpha\beta} n^\beta.
\end{align}

\vspace{10pt}
\section{The Neumann condition at first order}
\label{app:neumann_bc_at_first_order}
In the path integral, we restrict to configurations obeying the modified Neumann condition.
In particular, this means that perturbations satisfy
\begin{align}
   \delta( K_{ab} - h_{ab} K ) = T \delta h_{ab}.
\end{align}
The extrinsic curvature is given by
\begin{align}
    K_{\mu\nu} = h_\mu^\alpha h_\nu^\beta \nabla_\alpha n_\beta.
\end{align}
The variation of the normal is
\begin{align}
    \delta n_\mu = \frac 1 2 n_\mu n^\alpha n^\beta \delta g_{\alpha \beta}
\end{align}
and the variation of a Christoffel symbol can be written as
\begin{align}
    \delta \Gamma^\alpha_{\mu\nu} = \frac 1 2 g^{\alpha \beta} \left( \nabla_\mu \delta g_{\beta \nu} + \nabla_\nu \delta g_{\beta \mu} - \nabla_\beta \delta g_{\mu \nu} \right).
\end{align}
The variation of the induced metric, $h$, is
\begin{align}
    \delta h_{\mu\nu} &= \delta g_{\mu\nu} - n_\mu n_\nu n^\alpha n^\beta \delta g_{\alpha \beta} \\
    \delta h_\mu^\lambda &= h^{\lambda \nu} n_\mu n^\alpha \delta g_{\alpha \nu}\\
    \delta(h^{\mu\nu}) &= - \delta g^{\mu\alpha} h_\alpha^\nu - h^\mu_\beta \delta g^{\beta\nu} + \delta g^{\mu\nu} - n^\mu n^\nu \phi.
\end{align}
The second line can be shown by starting from $\delta(\delta^\mu_\nu) = \delta(n^\mu n_\nu - h^\mu_\nu)=0$.
With these formulas in hand, it is straightforward to show that
\begin{align}
    \delta K_{\mu\nu} = \frac 1 2 K_{\mu\nu} n^\rho n^\sigma \delta
  g_{\rho \sigma}+ n^\sigma \delta g_{\rho \sigma} ( n_\mu
  K^{\rho}_{\;\;\nu} + n_\nu  K^{\;\;\rho}_{\mu}) \notag \\
    - \frac 1 2 n^\rho \left( \nabla_\alpha \delta g_{\rho \beta}+\nabla_\beta \delta g_{\rho \alpha}-\nabla_\rho \delta g_{\alpha \beta} \right) h_\mu^\alpha h_\nu^\beta
\end{align}
and
\begin{align}
    \delta(h_{\mu\nu} K) = (\delta g_{\mu\nu} - n_\mu n_\nu n^\alpha
  n^\beta \delta g_{\alpha \beta}) K - h_{\mu\nu} K^{\alpha\beta}
  \delta g_{\alpha \beta} + \frac 1 2 h_{\mu\nu} K n^\rho n^\sigma
  \delta g_{\rho \sigma}  \notag \\
    - \frac 1 2 n^\rho h^{\alpha \beta} h_{\mu\nu} \left( \nabla_\alpha \delta g_{\rho \beta}+\nabla_\beta \delta g_{\rho \alpha}-\nabla_\rho \delta g_{\alpha \beta} \right).
\end{align}
The difference between these 
expressions is the variation in
boundary stress-energy, $T \delta h_{ab}$:
\begin{align}
    T (\delta g_{\mu\nu} - n_\mu n_\nu n^\alpha n^\beta \delta
  g_{\alpha \beta} ) = - \delta g_{\mu\nu} K + \left( \frac 1 2
  K_{\mu\nu} - \frac 1 2 h_{\mu\nu} K + n_\mu n_\nu K\right)n^\rho
  n^\sigma \delta g_{\rho\sigma}  \notag \\
    +n^\sigma \delta g_{\rho \sigma} ( n_\mu  K^{\rho}_{\;\;\nu} + n_\nu  K^{\;\;\rho}_{\mu}) \notag\\
    +h_{\mu\nu} K^{\alpha\beta} \delta g_{\alpha \beta}\notag\\
   + \frac 1 2 n^\rho \left( h^{\alpha \beta} h_{\mu\nu} -   h_\mu^\alpha h_\nu^\beta \right)\left( \nabla_\alpha \delta g_{\rho \beta}+\nabla_\beta \delta g_{\rho \alpha}-\nabla_\rho \delta g_{\alpha \beta} \right).
\end{align}
Metric perturbations $\delta g$ must satisfy these equations.
We can probe further by contracting with normal vectors and tangent
vectors.

\paragraph{Case 1.} Contracting both indices of the remaining terms
with the normal vectors, both sides automatically vanish. 

\paragraph{Case 2.} Contracting one index with the normal, and the
other with the projector onto the tangent space, we obtain 
\begin{align}
    0 = ( K^{\mu\rho} -  h^{\mu\rho} K - h^{\mu\rho} T )   h^\nu_\rho  \delta g_{\nu \sigma} n^\sigma.
\end{align}
This vanishes by the boundary condition.

\paragraph{Case 3.}
Finally, we consider projecting $\delta g$ onto the tangent space.
First, define
\begin{align}
    n^\rho n^\sigma \delta g_{\rho\sigma}   = \delta g_{nn} , \quad h^\alpha_\mu h^\beta_\nu \delta g_{\alpha \beta} = \delta g^\perp_{\mu\nu}.
\end{align}
Contracting both with $h^\mu_\nu$, we find
\begin{align}
    0 =  \frac 1 2  \left( K_{\mu\nu} - h_{\mu\nu} K \right)  \delta g_{nn}  
    + ( h_{\mu\nu} K^{\alpha\beta}- h^\alpha_\mu h^\beta_\nu (K + T) )
  \delta g^\perp_{\alpha \beta} \notag \\
    + \,\, \frac 1 2 n^\rho \left( h^{\alpha \beta} h_{\mu\nu} -   h_\mu^\alpha h_\nu^\beta \right)\left( \nabla_\alpha \delta g_{\rho \beta}+\nabla_\beta \delta g_{\rho \alpha}-\nabla_\rho \delta g_{\alpha \beta} \right).
\end{align}
This can be shortened using the variation formula for the Christoffel symbol:
\begin{align}
    0 =  \frac 1 2  \left( K_{\mu\nu} - h_{\mu\nu} K \right)  \delta g_{nn}  
    + ( h_{\mu\nu} K^{\alpha\beta}- h^\alpha_\mu h^\beta_\nu (K + T) ) \delta g^\perp_{\alpha \beta}
    + n_\rho \left( h^{\alpha \beta} h_{\mu\nu} -   h_\mu^\alpha h_\nu^\beta \right) \delta \Gamma^\rho_{\alpha\beta}.
\end{align}

\vspace{10pt}
\section{Covariance and boundaries}
At the heart of the covariant phase space formalism is the idea that a symmetry transformation can be represented by an action on the fields. However, the presence of a boundary makes the discussion slightly more difficult. 
In a fixed coordinate system, the action of a symmetry on the fields
cannot change the location of the boundary.
Thus, in order for the action to be invariant up to boundary terms, we
require that the symmetry reduces to a symmetry of the boundary at the
boundary itself. 
For example, diffeomorphism symmetries must not change the location to the boundary:
\begin{align}
\xi^\mu n_\mu \Big|_{\partial M} = 0.
\end{align}

One can introduce boundary quantities such as the normal vector in a
\emph{coordinate-invariant} way by using a function $f$, defined in
neighborhood of the boundary, which is smooth and negative except at
the boundary where it vanishes.
One can then define a (space-like) by
normal vector
\begin{align}
    n_\mu := \frac{\partial_\mu f}{\sqrt{\partial_\alpha f \partial_\beta f g^{\alpha\beta}}}.
\end{align}

However, the Lie derivative implementing a general diffeomorphism does
not act on $f$.
In order to preserve covariance, we therefore need the Lie derivative
\begin{align}
    \mathcal L_\xi n_\mu  = \xi^\nu \nabla_\nu n_\mu + n_\nu \nabla_\mu \xi^\nu
\end{align}
to agree with the transformation under a symmetry of fields on the boundary:
\begin{align}
    \delta_\xi n_\mu =  n_\mu n^\alpha n^\beta \nabla_\alpha \xi_\beta.
\end{align}
Thus, the allowed diffeomorphisms obey
\begin{align}
    0 = \Big( \xi^\nu \nabla_\nu n_\mu + n_\nu \nabla_\mu \xi^\nu -
  n_\mu n^\alpha n^\beta \nabla_\alpha \xi_\beta \Big) \Big|_{\partial
  M}.
\end{align}
The normal component of this equation vanishes and the only non-trivial content is obtained by projecting it onto the boundary,
\begin{align}
    0 = h^{\alpha\mu} \xi^\nu \nabla_\nu n_\mu + n_\nu h^{\alpha\mu} \nabla_\mu \xi^\nu.
\end{align}
From this it follows that
\begin{align}
    0 = h^{\alpha\mu} \partial_\mu (\xi^\nu n_\nu).
\end{align}
In fact, we can even be less restrictive by requiring that the above
equations do not hold in a neighborhood around the boundary, but at
the boundary to finite order in derivatives in normal direction.
For further discussion, see \cite{Harlow2019}.

\vspace{10pt}
\bibliographystyle{jhep}
\bibliography{refs.bib}

\end{document}